\documentclass[prl,twocolumn,showpacs,nobibnotes,floatfix,superscriptaddress]{revtex4}

\usepackage{graphicx,eucal}

\begin{document}


\title{Energetic Suppression of Decoherence in Exchange-Only Quantum Computation}
\author{Yaakov S. Weinstein}
\thanks{To whom correspondence should be addressed}
\email{weinstei@dave.nrl.navy.mil}
\author{C. Stephen Hellberg}
\email{hellberg@dave.nrl.navy.mil}
\affiliation{Center for Computational Materials Science, Naval Research Laboratory, Washington, DC 20375 \bigskip}

\begin{abstract}
Universal quantum computation requiring only the Heisenberg 
exchange interaction and suppressing decoherence via an energy gap is 
presented. The combination of an always-on exchange interaction between
the three physical qubits comprising the encoded qubit and a global magnetic 
field generates an energy gap between the subspace of interest and all other 
states. This energy gap suppresses decoherence. Always-on exchange couplings 
greatly simplify hardware specifications and the implementation of 
inter-logical-qubit gates. A controlled phase gate can be implemented using 
only three Heisenberg exchange operations all of which can be performed 
simultaneously. 
\end{abstract}

   
\maketitle

Encoding logical qubits (LQ) into subspaces of systems with more then one 
physical qubit is a powerful protocol in quantum information processing. 
Subspaces of this kind have been shown to protect quantum information from 
decoherence due to interactions with the environment, while still allowing 
for universal quantum computation \cite{Z1,Duan}. Experimental examples of 
these decoherence free subspaces (DFS) have been realized on nuclear magnetic 
resonance \cite{Evan1}, ion trap \cite{K}, and optical systems \cite{Alt}, 
have been suggested for superconducting qubits \cite{Zhou}, and have been 
used to implement encoded quantum algorithms \cite{Moh,Oller}. Encoded 
subspaces known as interaction free subspaces (IFS) can be used to protect 
information from always-on inter-qubit couplings. This reduces hardware 
constraints in certain system by circumventing the need to turn on and off 
two-qubit interactions. Rather, when the interaction is desired, the 
information is simply taken out of IFS \cite{ZZGF,BB}. A combination of DFS 
and IFS for universal computation has recently been proposed \cite{ZYZFG}. 
Finally, encoded subspaces allow performance of quantum logic maximizing the 
use of readily available operators while partially or completely avoiding 
operators that may add complexity to the computing hardware or a 
significant amount of time to the computation. Specifically, this type of 
subspace has been suggested to perform universal quantum computation with 
only the Heisenberg exchange interaction \cite{Div,Levy,Wu}.

Many proposals for the implementation of quantum computation, including 
quantum dots \cite{qdots} and other spin based methods \cite{Kane,ESR}, rely 
on the Heisenberg exchange as the means of inter-qubit interactions. This is, 
in part, due to the many desirable qualities found in Heisenberg exchange: 
it is short ranged, allows fast gate operation, and has an on-off ratio of 
many orders of magnitude \cite{BLD}. However, without encoding, the Heisenberg 
exchange alone is not universal. Experimental proposals of quantum computation 
thus supplement the Heisenberg exchange with one qubit gates which require 
local magnetic fields \cite{qdots} or g-factor engineering
\cite{Kato}. This puts strenuous demands on the quantum computing hardware 
and may significantly reduce computational speed. 

To circumvent the difficulty of one qubit gates in solid-state systems, 
methods have been devised to embed LQs in subspaces of more than 
one physical qubit in such a way that the exchange interaction becomes 
universal. One such encoding, introduced by DiVincenzo {\it et al} 
\cite{Div}, calls for each LQ to be embedded in three physical 
qubits. Exchange interactions within the LQ can perform global 
rotations while operations between LQs can be used for two-qubit gates. 
A minimization search found that 19 such operations were 
needed to perform a CNOT gate. 

A second encoding strategy to avoid single qubit rotations, suggested by Levy 
\cite{Levy,Benj1}, embeds the LQ into two physical qubits residing 
in inequivalent local environments. For example, if two quantum-dot qubits 
have different $g$-factors, a magnetic field will cause a phase difference 
between the physical qubits of the LQ and modulation of the 
exchange coupling at the qubit Rabi frequency performs $z$-rotations. 
Heisenberg exchange between the two physical qubits of the LQ 
perform an $x$-rotation, and two-qubit gates are performed via exchange 
coupling between qubits of different LQs. Using this scheme a 
controlled phase shift gate can be achieved with two exchange couplings and 
two $z$-rotations.

\begin{figure}
\includegraphics[height=5.8cm]{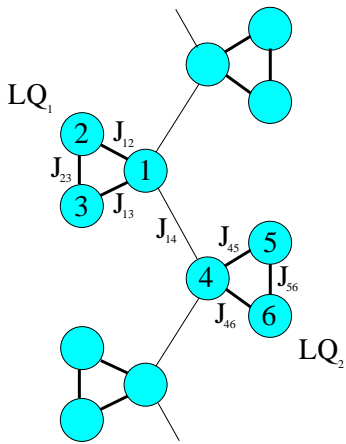}
\caption{\label{Dots}
(Color online) Suggested layout for encoded or logical qubits (LQ). Each 
LQ is composed of three physical qubits arranged in an
equilateral triangle. The lines between physical qubits represent exchange 
interactions. The thick lines within LQs show always-on exchange 
interactions which are modified to perform single qubit rotations.
The thin lines between LQs represent weaker exchange interactions
which are turned on to perform two-qubit gates.
}
\end{figure}

In this Letter we present an alternative scheme for exchange-interaction-only 
universal computation. Our scheme is similar to that of DiVincenzo {\it et al}
\cite{Div} in that the LQ is encoded in the $S_z = +1/2$ subspace 
of a three qubit system. Single qubit rotations are then performed via the 
exchange coupling between physical qubits within the LQ and no 
$g$-factor engineering is required. In our scheme the LQ in 
its idle mode has equal and constant exchange couplings between the three 
physical qubits, analogous to previously proposed IFS architectures with 
always-on qubit couplings \cite{ZZGF,BB}. To generate three equal couplings 
between the physical qubits, we propose arranging them in equilateral 
triangles as shown in Fig. \ref{Dots}. In addition, the system is placed 
in a uniform magnetic field. The combination of the always-on couplings and 
the magnetic field creates an energy gap between the LQ subspace, which 
is the degenerate ground state, and the other states of the system. This gap 
makes the system more robust against decoherence \cite{superco}. Single 
qubit rotations are still performed by changing the strength of the exchange
coupling between two physical qubits and can now be done both in 
the positive and negative sense. Two LQ operations, however, 
are simplified enormously due to the gap. With adiabatic modification of 
the interactions the system returns to the logical subspace when the 
couplings between LQs are turned off. Adiabatic evolution of interactions 
is already a requirement for {\em all} approaches using spins in quantum 
dots \cite{SLM}. A conditional phase gate requires modification of only 
three exchange couplings all of which can be performed simultaneously. 

We now analyze the dynamics of the above system and outline how to implement 
a universal quantum computation. The Heisenberg exchange interaction 
between two spins is
\begin{equation}
H_{ij} = J_{ij}{\bf S}^i\cdot{\bf S}^j = J_{ij}\left(S_x^iS_x^j+S_y^iS_y^j+S_z^iS_z^j\right),
\end{equation}
where ${\bf S} = {\bf \sigma}/2$ and $J_{ij}$ is the coupling strength between
spins $i$ and $j$. When placed in a uniform magnetic field there is an 
additional Zeeman interaction
\begin{equation}
H_z = -h\sum_i S_z^i
\end{equation}
where $h = gB$ depends on the gyromagnetic ratio, $g$, which is the same for 
all dots, and the magnetic field strength, $B$. The Hamiltonian of a single 
LQ is thus
\begin{equation}
H_{LQ} = H_z+H_{12}+H_{13}+H_{23}.
\end{equation}

\begin{figure}
\includegraphics[height=5.8cm, width=8cm]{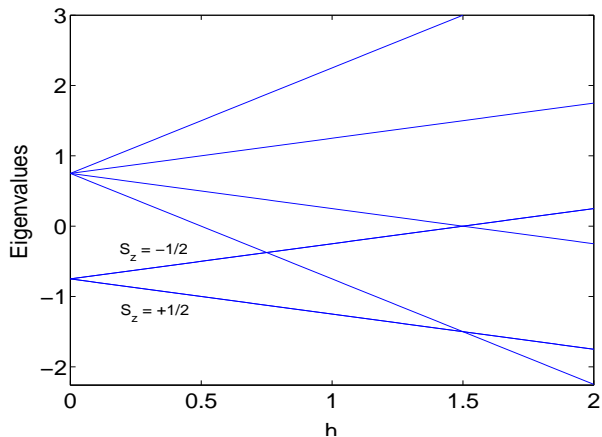}
\caption{\label{OneLQ}
(Color online) Eigenvalues of one LQ as a function of global 
magnetic field, $h = gB$. The eigenvalues marked $S_z = +1/2$ and 
$S_z = -1/2$ are doubly degenerate. For suppression of decoherence, 
maximizing the gap between the $S_z = +1/2$, states used for the LQ 
and the remaining states is desirable. The gap is maximized for 
$h = .75$. 
}
\end{figure}

In the LQ's idle mode, the coupling exchange between all pairs 
of physical qubits is equal (to 1 for simplicity) 
$J_{12} = J_{13} = J_{23} = 1$. When $h$ is non-zero the system has a 
doubly degenerate ground state corresponding to the $S_z = +1/2$ subspace, 
as seen in Fig. \ref{OneLQ}. The orthogonal basis states of the LQ are 
chosen to span this space. In terms of the physical qubits the states are:
\begin{eqnarray}
|0_L\rangle &=& \frac{1}{\sqrt{2}}|\uparrow\uparrow\downarrow\rangle-\frac{1}{\sqrt{2}}|\uparrow\downarrow\uparrow\rangle \nonumber\\
|1_L\rangle &=& \frac{1}{\sqrt{6}}|\uparrow\uparrow\downarrow\rangle + \frac{1}{\sqrt{6}}|\uparrow\downarrow\uparrow\rangle - \frac{2}{\sqrt{6}}|\downarrow\uparrow\uparrow\rangle.
\end{eqnarray}
With this choice, initialization can be done by breaking the degeneracy 
through raising or lowering one of the couplings and waiting for the system 
to decay to its ground state \cite{Div}. Also note that $|0_L\rangle$ 
incorporates the singlet state between spins 2 and 3 and $|1_L\rangle$ 
incorporates triplet states between the same spins. State $|0_L\rangle$ is 
antisymmetric with respect to exchange of sites 2 and 3, while state 
$|1_L\rangle$ is symmetric for this exchange. This allows for read-out 
of the LQs via singlet-triplet measurement schemes \cite{Kane}. 

We choose the strength of the external magnetic field so as to maximize the 
energy gap between the doubly degenerate ground state and other states. The
energy gap protects against many types of decoherence by forcing these
processes to add energy to the system \cite{superco}. For the above encoding 
single-spin flips and arbitrary global magnetic fields cannot effect the 
LQs without forcing the system to leave the LQ subspace, requiring energy to 
overcome the gap. Single-spin phase operations, are not protected by the gap 
since they do not force the system out of the subspace and, thus, do not 
require energy. In general, the energy gap increases the 
robustness of the computation by inhibiting decohering processes which would 
cause the system to leave the subspace. The optimal external magnetic field 
to maximize the energy gap is $h = .75$ as shown in Fig. \ref{OneLQ}.

\begin{figure}
\includegraphics[height=5.8cm, width=8cm]{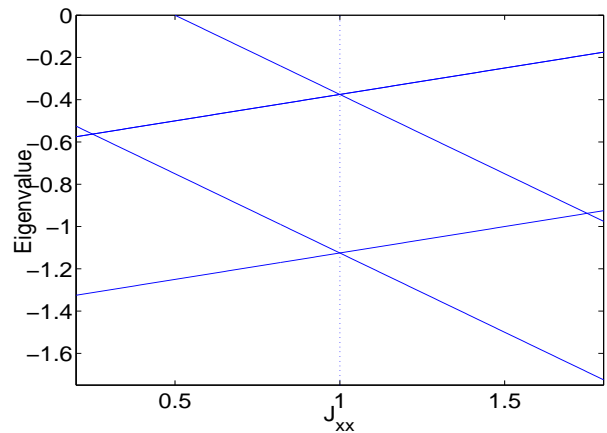}
\caption{\label{gates}
(Color online) Eigenvalues as a function of coupling strength 
$J_{xx} = J_{12,13,23}$ during application of one-LQ gates. The idle-mode
exchange coupling value, $J_{xx} = 1$, is marked by the vertical dashed line. 
A $\sigma_z$ rotation is implemented by changing $J_{23}$ from its 
idle mode value. This breaks the degeneracy of the $|0_L\rangle$, 
$|1_L\rangle$ ground state but does not mix the states. Changing $J_{12}$ 
or $J_{13}$ implements a rotation around an axis in the $x-z$ plane 
$120^{\circ}$ away from the $z$-axis by breaking the doubly degenerate 
ground state into the states 
$\frac{\sqrt{3}}{2}|0_L\rangle-\frac{1}{2}|1_L\rangle$ 
and $\frac{1}{2}|0_L\rangle+\frac{\sqrt{3}}{2}|1_L\rangle$. Combinations of 
the two rotations can implement arbitrary one-qubit rotations. 
The energy gap to leave the LQ subspace remains
as long as the strength of the coupling being changed
does not cause a level crossing; that is, $0.25 < J_{xx} < 1.75$.
}
\end{figure}

Changing the strength of exchange couplings between physical qubits within 
the LQ, as in \cite{Div}, performs single LQ rotations while always keeping 
the state of the system in the LQ subspace. Because the LQs have non-zero 
idle mode exchange couplings, positive and negative LQ rotations can be 
performed. 
\begin{eqnarray}
H_1 &=&
\frac{1}{4}
\left(
\begin{array}{cc}
-J_{12}+2J_{23}-J_{13} & \sqrt{3}(J_{12}-J_{13}) \\
\sqrt{3}(J_{12}-J_{13}) &  J_{12}-2J_{23}+J_{13}\\
\end{array}
\right)
\end{eqnarray}
is the LQ Hamiltonian for the general case written in the basis 
$[|0_L\rangle,|1_L\rangle]$. The Hamiltonian is defined within a constant 
times the identity which merely applies a global phase. From the general 
Hamiltonian we see that changing only $J_{23}$ performs a $z$-rotation and 
changing only $J_{12}$ or $J_{13}$ performs a rotation about the axis in 
the $x-z$ plane $120^{\circ}$ from the $z$ axis. The eigenvalues of a LQ for 
these operations as a function of the change in coupling is shown 
in Fig. \ref{gates}. Combinations of the above are sufficient to perform 
arbitrary $SU(2)$ rotations. We also note that it is possible to achieve 
$x$-rotations by changing $J_{12}$ and $J_{23}$ such that 
$\Delta J_{12} = 2\Delta J_{23}$.

\begin{figure}
\includegraphics[height=5.8cm, width=8cm]{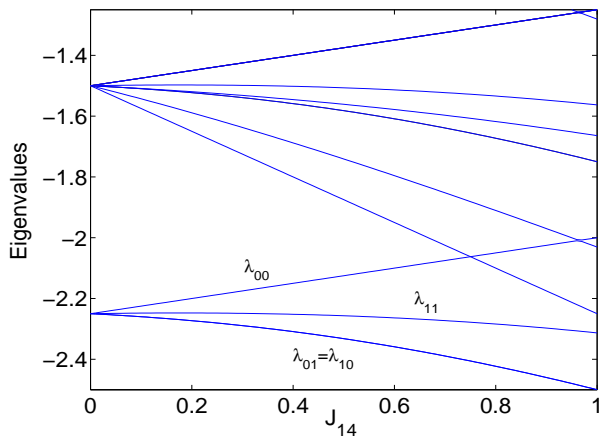}
\caption{\label{TwoLQ}
(Color online) Eigenvalues of two LQs as a function of $J_{14}$.
The coupling breaks the LQ's four-fold degenerate ground state 
into the states (from top to bottom) $|00_L\rangle$, $|11_L\rangle$, and 
the degenerate $|01_L\rangle$ and $|10_L\rangle$. The energy gap to leave the 
LQ subspace remains for all $J_{14} < .75$. 
}
\end{figure}

There are numerous ways to couple LQs of the type described to allow 
universal quantum computation. Here, we describe one of the 
simplest approaches, a single exchange coupling between physical qubits 1 
and 4 of Fig. \ref{Dots}. This coupling keeps the symmetry of the individual 
LQs, namely exchange of spins 2 and 3. Thus, a $J_{14}$ couplings 
results in a diagonal inter-LQ interaction in the two-LQ basis 
$[|00_L\rangle,|01_L\rangle,|10_L\rangle,|11_L\rangle]$. We can 
write the effective Hamiltonian as
\begin{eqnarray}
H_2 &=&
\left(
\begin{array}{cccc}
\lambda_{00} & 0 & 0 & 0 \\
0 & \lambda_{01} & 0 & 0 \\
0 & 0 & \lambda_{01} & 0 \\
0 & 0 & 0 & \lambda_{11} \\
\end{array}
\right),
\end{eqnarray}
where $\lambda_{10} = \lambda_{01}$. The eigenvalues as a function of $J_{14}$ 
are plotted in Fig. \ref{TwoLQ} along with the low-lying states outside of the 
LQ. The energy gap for leaving the logical subspace remains for all 
$J_{14} < 0.75$. The eigenvalues themselves can be solved from the 
lowest solutions to the following equations:
\begin{eqnarray}
0 &=& 4\lambda_{00} - (J_{14}-3) \nonumber\\
0 &=& 16\lambda_{11}^2 + 8J_{14}\lambda_{11} - 3J_{14}^2 + 16J_{14} + 27 \\
0 &=& 64\lambda_{01}^3+16(J_{14}+9)\lambda_{01}^2-4(5J_{14}^2-14J_{14}+9)\lambda_{01}\nonumber\\
  & & +3J_{14}^3-23J_{14}^2+37J_{14}-81 \nonumber.
\end {eqnarray}
The solutions, particularly for $\lambda_{01}$, are rather involved and are
omitted. The Taylor expansions for small $J_{14}$ are
\begin{eqnarray}
\lambda_{00} &=& -\frac{9}{4}+\frac{1}{4}J_{14} \nonumber\\
\lambda_{11} &=& -\frac{9}{4}-\frac{1}{12}J_{14}-\frac{4}{27}J_{14}^2-\frac{8}{243}J_{14}^3+\cdots \\
\lambda_{01} &=& -\frac{9}{4}+\frac{1}{36}J_{14}-\frac{2}{27}J_{14}^2-\frac{110}{6561}J_{14}^3+\cdots. \nonumber
\end{eqnarray}

When the interaction is modified adiabatically \cite{SLM}, the system is left
in the logical subspace after $J_{14}$ is turned off. This is due to the 
energy gap generated by the always-on exchange interactions within the LQ.
Removing leakage outside the logical subspace greatly simplifies the gate
compared with the gate required in the case with no energy gap \cite{Div}.

The two-LQ interaction plus equal $z$ rotations on each of the 
two LQs performs a conditional phase gate. Thus, we can implement a 
universal set of gates \cite{Bar}: single qubit rotations, by 
changing the exchange couplings within a LQ, and conditional phase 
gates via the two LQ interaction, turning on the $J_{14}$ 
exchange coupling. We further note that the operations to perform the 
conditional phase gate may be done simultaneously, thus the gate requires 
only \emph{one} time interval.

Up to this point we have not specified a physical system for our scheme. 
We now do so to insure that the strength of magnetic field required is 
reasonable. Of the materials proposed for quantum computation with quantum 
dots \cite{L2,WI,Marcus,Kouwen}, GaAs has the $g$-factor with smallest 
magnitude, $g = -.44$. Approximating the required strength of the magnetic
field using values found in Ref. \cite{DL}, $J \sim 7 \;\mu$eV 
\cite{calc}, the optimal value of $h = .75$ in units of $J$ can be achieved 
with magnetic field strength $B \sim .2\;$T. For a LQ in the idle 
mode, this gives an energy gap between the LQ subspace and the rest 
of the states of $\Delta E \sim 5\;\mu$eV. In Si, which has a larger 
$g$-factor, the required magnetic field is even smaller.

In conclusion, we have demonstrated a scheme which suppresses 
decoherence via an energy gap between the logical qubit states and all
other states and performs universal quantum computation using only the 
Heisenberg exchange process. The scheme calls for embedding each logical 
qubit in three physical qubits coupled by always-on Heisenberg exchange
interactions. Positive and negative single qubit rotations are simply 
implemented by changing exchange interactions between physical qubits 
within a logical qubit. Placing the system in a magnetic field opens an 
energy gap between the logical qubit subspace and all other states. The energy 
gap greatly simplifies the implementation of the two logical qubit gates 
protects against decoherence. Conditional phase gates can be implemented with 
just three exchange gates all of which can be done simultaneously. The 
conditional phase gate combined with the one logical qubit gates are 
sufficient for universal quantum computation.


\begin{thebibliography}{99}
\bibitem{calc}
Ref. \cite{DL} provides experimental estimates for the single-dot charging 
energy, $E_C \sim 500\;\mu$eV, and tunnel coupling, $t \sim 30\;\mu$eV.
The coupling between dots is given by $J = 4t^2/E_C \sim 7\;\mu$eV.
\end{thebibliography}
\end{document}